\documentclass[conference]{IEEEtran}
 \IEEEoverridecommandlockouts

\usepackage[cmex10]{amsmath}
\usepackage{amssymb}
\usepackage{epsfig,epsf,pstricks,pgf}
\usepackage{cite}

\usepackage{enumerate}

\usepackage{amsthm}
\usepackage{psfrag}

\usepackage{times}
\usepackage{epsfig}
\usepackage{amsmath}
\usepackage{amsfonts}
\usepackage{algorithm}
\usepackage{algorithmic}
\usepackage{graphicx}
\usepackage{amssymb}
\usepackage{amstext}
\usepackage{latexsym}
\usepackage{color}
\usepackage{ifthen}
\usepackage{multirow}
\usepackage{verbatim}
\usepackage{array,tabularx}
\usepackage{epsf,pstricks,pgf,pst-node}
\usepackage{color}
\usepackage{url}

\newcommand{\mbf}{\mathbf}
\newcommand{\mcl}{\mathcal}

\newcommand{\und}{\underline}

\theoremstyle{definition}
\newtheorem{definition}{Definition}
\newtheorem{example}{Example}

\IEEEoverridecommandlockouts
\begin{document}

% paper title
%\title{Deterministic Algorithm for the Cooperative Data Exchange Problem}
\title{Data Exchange Problem with Helpers }
\author{% aefraer \IEEEauthorrefmark{baer}
\IEEEauthorblockN{Nebojsa Milosavljevic, Sameer Pawar, Salim El Rouayheb, Michael Gastpar\IEEEauthorrefmark{2}\thanks{\IEEEauthorrefmark{2}Also with the School of Computer and Communication Sciences, EPFL, Lausanne, Switzerland.} and Kannan Ramchandran}
\IEEEauthorblockA{Department of Electrical Engineering and Computer Sciences \\
University of California, Berkeley \\
%%Berkeley, CA, 94720, USA \\
Email: \{nebojsa, spawar, salim, gastpar, kannanr\}@eecs.berkeley.edu}
\thanks{This research was funded by the NSF grants
(CCF-0964018, CCF-0830788), a DTRA grant (HDTRA1-09-1-0032), and in part by an
AFOSR grant (FA9550-09-1-0120).}
%}
}
%\thanks{hello where am I??}

% make the title area
\maketitle

%\linespread{0.97}

%!TEX root = isit2012.tex
\begin{abstract}
%In this paper we construct a deterministic polynomial time algorithm that achieves the minimum linear cost for a data exchange problem with helpers. We assume that a group of terminals individually have partial information of a common file, but collectively have sufficient information to recover the complete file. The subset of terminals that are interested in recovering the file are called {\em users} while the compliment set of terminals are called {\em helpers}. The goal is to determine transmission rates of all the terminals such that the user terminals recover the file and the weighted number of bits that these terminals need
%to exchange over a noiseless public channel in order achieve this goal. Using established connections to the  multi-terminal secrecy problem,
%our algorithm also implies  a polynomial-time method for constructing a maximum size secret shared key in the presence of an eavesdropper. We consider the
%following types of side-information settings: (i) side information in the form  of uncoded packets of the
%file, where the users' side-information consists of subsets of the file;  (ii) side information in the  form of
%linearly correlated packets, where the users have access to linear combinations of the file packets; and (iii) the
%general setting where the the users' side information has an arbitrary (i.i.d.) correlation structure.
%We provide a polynomial-time algorithm (in the number of users) that
%finds the optimal rate allocations among these users,  then determines an explicit  optimal transmission scheme for cases (i) and (ii).
In this paper we construct a deterministic polynomial time algorithm for the problem where
a set of users is interested in gaining access to a common file, but where each has only partial knowledge of the file.
We further assume the existence of another set of terminals in the system, called helpers, who are not interested
in the common file, but who are willing to help the users.
Given that the collective information of all the terminals is sufficient to allow
recovery of the entire file, the goal is to minimize the (weighted) sum of bits that these terminals need
to exchange over a noiseless public channel in order achieve this goal. Based on established connections to the  multi-terminal secrecy problem,
our algorithm also implies  a polynomial-time method for constructing the largest shared secret key in the presence of an eavesdropper. We consider the
following side-information settings: (i) side-information in the form  of uncoded packets of the
file, where the terminals' side-information consists of subsets of the file;  (ii) side-information in the  form of
linearly correlated packets, where the terminals have access to linear combinations of the file packets; and (iii) the
general setting where the the terminals' side-information has an arbitrary (i.i.d.) correlation structure.
We provide a polynomial-time algorithm (in the number of terminals) that
finds the optimal rate allocations for these terminals, and then determines an explicit  optimal transmission scheme for cases (i) and (ii).
\end{abstract} 
%!TEX root =isit2012.tex
\section{Introduction}\label{sec:intro}
In recent years cellular systems  have witnessed significant improvements in terms of data rates, and are nearly
approaching the theoretical limits in terms of the physical layer spectral efficiency. At the same time, the rapid growth in
the popularity of data-enabled mobile devices, such as smart phones and tablets,
and the resulting explosion in demand for more throughput are challenging our abilities
even with the current highly efficient cellular systems. One of the major bottlenecks in scaling the throughput with the
increasing number of mobile devices is the ``last mile'' wireless link between the base station and the mobile devices -- a
resource that is shared among many terminals served within the cell. This motivates the study of paradigms where cell phone
devices can  cooperate among themselves to get the desired data in a peer-to-peer fashion without solely relying on the base station.

\begin{figure}
\begin{center}
\psset{unit=0.40mm}
\begin{pspicture}(0,0)(170,90)
\small{
\psframe(60,10)(110,30)\rput(85,20){\small{Base Station}}
\rput(85,2){$\{w_1,w_2,w_3,w_4\}$}
\rput(-6,20){$\left\{
                \begin{array}{c}
                  w_2 \\
                  w_3 \\
                \end{array}
              \right\}$}
\psframe(12,15)(37,25)
\rput(25,20){user $1$}
\psframe(133,15)(159,25)
\rput(146,20){user $2$}
\rput(177,20){$\left\{
                 \begin{array}{c}
                   w_1 \\
                   w_2 \\
                   w_4 \\
                 \end{array}
               \right\}
$}
\rput(85,80){$\left\{
                \begin{array}{c}
                  w_1 \\
                  w_3 \\
                \end{array}
              \right\}
$}
\psframe(72,52)(98,64)
\rput(85,58){helper}
\psline{->}(60,20)(37,20)
\psline{->}(110,20)(133,20)
\psline{->}(85,30)(85,52)
}
\end{pspicture}
\end{center}
\caption{An example of the data exchange problem with helpers. A base station has a file formed of four packets $w_1,\dots, w_4\in
\mathbb{F}_{q^n}$ and wants to  deliver it to two users over an unreliable wireless channel.
Additionally, there is a
terminal in the system that is in the range of the base station, but he is not interested in the file.
However, he is willing to help the two users to obtain the file. The base station
stops transmitting once all terminals collectively have all the packets, even if individually they have only subsets of the
packets. They can then cooperate among themselves to recover the users' missing packets.
If the goal is to minimize the total number of communicated bits, helper transmits packet
$w_1+w_3$, while user~$2$ transmits packet $w_4$, where the addition is in the  field $\mathbb{F}_{q^n}$.}
\label{fig:model_raw}
\end{figure}
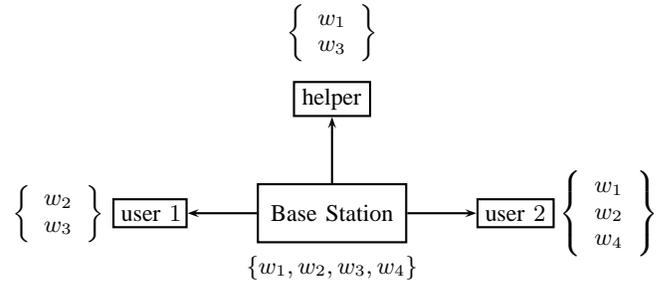

An example of such a setting is shown in Figure~\ref{fig:model_raw}, where a base station
wants to deliver the same file to multiple  geographically-close users over an unreliable wireless downlink. We assume
that some terminals, which are in the range of the base station, are not interested in the file, but due to their proximity to the base station,
they are able to overhear some of its transmissions. Moreover, we assume that these terminals are willing to help in distributing the file to
the respective users. We will refer to these terminals as \emph{helpers}.
%Due to their altruistic nature, we call them helpers.
In the example of Figure~\ref{fig:model_raw} we assume that the file
consists of four equally sized packets $w_1$, $w_2$, $w_3$ and $w_4$ belonging to some finite field $\mathbb{F}_q^n$.
Suppose that after a few initial transmission attempts by the base station, the three terminals (including one helper) individually receive only
parts of the file (see Figure~\ref{fig:model_raw}), but collectively have the entire file. Now, if all terminals
are in close vicinity  and can communicate with each other, then, it is much more desirable and efficient, in terms
of resource usage, to reconcile the file among users by letting all terminals ``talk'' to each other without involving the  base station.
The cooperation among the terminals has the following advantages:
\begin{itemize}
\item Local communication among terminals has a smaller footprint in terms of interference, thus allowing one to use the shared resources (code, time or frequency)
freely without penalizing the base station's resources, \emph{i.e.}, higher resource reuse factor.
\item Transmissions within the close group of terminals is much more reliable than from the base station to any terminal due to geographical proximity of terminals.
\item This cooperation allows for the file recovery even when the connection to the base station is either unavailable after the initial phase of transmission, or it is too weak to meet the delay requirement.
\end{itemize}

The problem of reconciling a file among
multiple  wireless users having parts of it while minimizing  the  cost in terms of the total number of bits exchanged
is known in the literature as the {\em data exchange problem} and was introduced by El Rouayheb {\em et al.} in
\cite{SSS10}. In the problem formulation of the data exchange problem it is assumed that all the terminals in the system
are interested in recovering the entire file, \emph{i.e.}, there are no helpers. For data exchange problem without helpers a randomized algorithm was proposed in \cite{SSBE10} and \cite{ozgul2011algorithm}, while a deterministic polynomial time algorithms was proposed in \cite{CXW10}, \cite{tajbakhshmodel}.
%In a more general setting of the problem, one can consider  minimizing a different cost function, a
%``weighted sum rate'', to accommodate the scenario when transmissions from different terminals have different
%costs. This problem was studied by Ozgul {\em et al.} \cite{ozgul2011algorithm}, where the authors proposed a
%randomized algorithm that achieves this goal with {\em high probability} provided that the underlying field size is
%large enough. In~\cite{courtade2011} and~\cite{milosavljevic2011optimal} the authors proposed a deterministic polynomial time algorithm
%that solves the original data exchange problem with arbitrary linear cost.

In this paper we consider a scenario with helpers, and linear communication cost.
W.r.t. the example considered here, if user~$1$, user~$2$ and the helper transmit $R_1,R_2$ and $R_3$ bits, respectively,
the data exchange problem with helpers would correspond to minimizing the weighted sum-rate $\alpha_1R_1+\alpha_2R_2+\alpha_3R_3$ such that,  when
the communication is over, user~$1$ and user~$2$ can recover the entire file. It can be shown that
for the case when $\alpha_1=\alpha_2=\alpha_3=1$, the minimum communication cost is $2$ and can be achieved by the following coding scheme: user $2$
transmits packet $w_4$, and the helper transmits $w_1+w_3$, where the addition is over the
underlying field $\mathbb{F}_{q^n}$. This corresponds to the optimal rate allocation $R_2=R_3=1$ symbol in
$\mathbb{F}_{q^n}$. If there was no helper in the system, it would take a total of $3$ transmissions to reconcile the
file among the two users. That is user~$1$ has to transmit $w_3$ and user~$2$ transmits $w_1$ and $w_4$. Thus, the  helpers can contribute to lowering the total communication cost in the system.

The discussion above considers  only a simple form of side-information, where different terminals observe partial uncoded
``raw'' packets  of the original file. Content distribution networks are increasingly using coding, such as Fountain
codes or linear network codes, to improve the system efficiency~\cite{luby2002lt}. In such  scenarios, the side-information representing
the partial knowledge gained by the terminals would be coded and in the form of linear combinations of the original file packets, rather than the raw packets themselves. The previous two cases of side-information (``raw'' and
coded) can be regarded as special cases of the more general problem where the side-information has arbitrary
correlation among the data observed by the  different terminals and where the goal is to minimize the weighted total
communication cost.
In~\cite{CN04} Csisz\'ar and Narayan posed a related security
problem referred to as  the  ``multi-terminal key agreement'' problem. They showed  that obtaining the file among the  users
in minimum number of bits exchanged over the public channel is sufficient to maximize the size of the secret key shared between the users. This result establishes  a connection between the Multi-party key agreement and the Data exchange problem with helpers. \cite{CN04} solves the key agreement
problem by formulating it as a linear program (LP) with an exponential number of rate-constraints, corresponding to all
possible cut-sets that need to be satisfied.

%In~\cite{milosavljevic2011optimal} we proposed a deterministic polynomial time algorithm that solves the data
%exchange problem with linear cost function and arbitrary correlated side information, when all the users in the system
%are interested in achieving omniscience.

%In this paper, we propose a \emph{deterministic polynomial time} algorithm for finding an optimal rate allocation, w.r.t.
%a linear weighted sum-rate cost, that solves the data exchange problem with helpers for an arbitrarily correlated side information.

In this paper, we  make the following contributions. First,  we provide a \emph{deterministic polynomial time}
algorithm for finding an optimal rate allocation, w.r.t. a linear weighted sum-rate cost needed to
deliver the file to all users when all terminals have arbitrarily correlated side-information. For the  data exchange problem with helpers,  this
algorithm computes the optimal rate allocation in polynomial time  for the case of  linearly coded side-information
(including the ``raw'' packets case) and for the general linear cost functions (including the sum-rate case).
Second, for the the data exchange problem with helpers, with raw or linearly coded
side-information, we propose an efficient communication scheme design based on the algebraic network coding framework~\cite{KM03}, \cite{H05}.

%!TEX root = Salim_DatExc.tex
\section{System Model and Preliminaries}\label{sec:model}
In this paper, we consider a set up with $m$ terminals out of which some subset of them is interested
in gaining access to a file or a random process. Let $X_1,X_2,\ldots,X_m$, $m \geq 2,$ denote the components of a discrete
memoryless multiple source (DMMS) with a given joint probability mass function. Each user $i \in
\mcl{M}\triangleq \{1,2,\ldots,m\}$ observes $n$ i.i.d. realizations of the corresponding random variable $X_i$.

Let $\mcl{A}=\{1,2,\ldots,k\}\subseteq \mcl{M}$ be the subset of terminals, called users, who are interested in gaining access to the file,
\emph{i.e.}, learning the joint process $X_{\mcl{M}}=(X_1,\ldots,X_m)$. The remaining terminals $\{k+1,\ldots,m\}$ serve as helpers,
\emph{i.e.}, they are not interested in recovering the file, but they are willing to help users in the set $\mcl{A}$ to obtain it.
In \cite{CN04}, Csisz\'ar and Narayan showed that deliver the file to all users in a setup with
general DMMS {\em interactive communication is not needed}. As a
result, in the sequel WLOG we can assume that the transmission of each user is only a function of its own initial
observations. Let $F_i \triangleq f_i(X^n_i)$ represent the transmission of the user $i \in \mcl{M}$, where $f_i(\cdot)$ is any
desired mapping of the observations $X^n_i$. For each user in $\mcl{A}$ in order to recover the entire file, transmissions $F_i$, $i\in
\mcl{M}$, should satisfy,
\begin{align}
\lim_{n \rightarrow \infty} \frac{1}{n} H(X_{\mcl{M}}^n|\mbf{F},X_{t_l}^n) = 0,~~~\forall t_l \in \mcl{A}, \label{eq:decode}
\end{align}
where $X_{\mcl{M}}=(X_1,X_2,\ldots,X_m)$.
\begin{definition}
A rate tuple $\mbf{R}=(R_1,R_2,\ldots,R_m)$ is an {\em achievable data exchange (DE) rate tuple} if there exists a communication scheme with transmitted messages $\mbf{F}=(F_1,F_2,\ldots,F_m)$ that satisfies \eqref{eq:decode}, and is such that
\begin{align}
R_i = \lim_{n \rightarrow \infty} \frac{1}{n} H(F_i),~~~\forall i \in \mcl{M}.
\end{align}
\end{definition}

% The goal in this paper is to devise a scheme of polynomial complexity for achieving the omniscience at each terminal such that the appropriately defined {\em communication cost} is minimized. In the sequel, we consider the {\em linear function} of the rates as an objective cost function.
It is easy to show using cut-set bounds that all the achievable \emph{DE} rate tuple's necessarily belong to the following region
\begin{align}
\mcl{R}\triangleq \left\{\mbf{R}: R(\mcl{S})\geq H(X_{\mcl{S}}|X_{\mcl{S}^c}),~\forall \mcl{S}\subset \mcl{M},~\mcl{A} \not \subseteq \mcl{S}\right\}, \label{cut_set}
\end{align}
where $R(\mcl{S}) = \sum_{i \in \mcl{S}} R_i$. Also, using a random coding argument,  it can be shown that the rate region $\mcl{R}$ is an achievable rate region \cite{CN04}.

In this work, we aim to design a polynomial complexity algorithm that delivers the file to all users in $\mcl{A}$ while simultaneously minimizing
a linear communication cost function $\sum_{i=1}^m \alpha_i R_i$, where $\und{\alpha} \triangleq (\alpha_1,\cdots,\alpha_m), 0 \leq \alpha_i < \infty$,
is an $m-$dimensional vector of non-negative finite weights. We allow $\alpha_i$'s to be arbitrary non-negative constants, to account for the case when communication
of some group of terminals is more expensive compared to the others, \emph{e.g.}, setting $\alpha_1$ to be a large value compared to the other weights minimizes the rate allocated to the user $1$. This goal can be formulated as the following linear program:
%\begin{align}
%\min_{\mbf{R}} \sum_{i=1}^m \alpha_i R_i,~~~~\text{s.t.}~~~\mbf{R} \in \mcl{R}, \label{problem1}
%\end{align}
%where $\mcl{R}$ is defined in~\eqref{cut_set}.
\begin{align}
&\min_{\mbf{R}} \sum_{i=1}^m \alpha_i R_i, \label{problem1} \\
&~~~~~\text{s.t.}~~R(\mcl{S})\geq H(X_{\mcl{S}}|X_{\mcl{S}^c}),~\forall \mcl{S}\subset \mcl{M},~\mcl{A} \not \subseteq \mcl{S}. \nonumber
\end{align}

\subsection{Finite Linear Source Model}
In general an efficient content distribution networks use coding such as fountain codes or linear network codes.
This results in terminals' observations to be in the form of linear combinations of the original packets forming the file, rather
than the uncoded data themselves as is the case in conventional `Data Exchange problem'. This linear correlation source model is known in literature as {\em Finite linear source} \cite{CZ10}.

Next, we briefly describe the finite linear source model. Let $q$ be some power of a prime.
Consider the $N$-dimensional random vector $\mathbf{W} \in \mathbb{F}^N_{q^n}$
whose components are independent and uniformly distributed over the elements of $\mathbb{F}_{q^n}.$
Then, in the linear source model, the observation of $i^{th}$ user is simply given by
\begin{align}
\mathbf{X}_{i} = \mathbf{A}_i \mbf{W}, \ i \in \mcl{M}, \label{model:eq1}
\end{align}
where $\mathbf{A}_i \in \mathbb{F}_{q}^{\ell_i \times N}$ is an observation matrix
%\footnote{ The entries in the
%observation matrix $A_i, \forall i \in {\cal M}$ denote the coefficients of the code, e.g., Fountain code or linear
%network code, used by the base station and hence belong to the smaller field $\mathbb{F}_q$ rather than the field
%$\mathbb{F}_{q^n}$ to which the data packets belong. This assumption is justified since the coding coefficients are typically stored in the packet in an overhead of size
%negligible compared to the packet length.
%It also, allows one to ``split'' the packets in order to achieve rational rates.
%}
for the user~$i$.

It is easy to verify that for the finite linear source model,
\begin{align}
\frac{H(X_i)}{\log q^n} = \text{rank}(\mbf{A}_i). \label{rank_entropy}
\end{align}
Henceforth for the finite linear source model we will use the entropy of the
observations and the rank of the observation matrix interchangeably. 
\section{Deterministic Algorithm} \label{sec:alg}

We begin this section by exploring the case when the set $\mcl{A}$ consists of only one user.
Then, by using the methodology of~\cite{lun2006minimum}, we extend our
solution to the case when the set $\mcl{A}$ has arbitrary number of users.

\subsection{Deterministic Algorithm when $|\mcl{A}|=1$}

Let the user $t_1 \in \mcl{M}$ be the only one user interested in a file, \emph{i.e.}, $\mcl{A}=\{t_1\}$. This is known as a multi-terminal Slepian-Wolf
problem~\cite{CT06} for which the achievable rate region has the following form:
\begin{align}
\mcl{R}_1 = \left\{ \mbf{R} : R(\mcl{S})\geq H(X_{\mcl{S}}|X_{\mcl{S}^c},X_1),~\forall \mcl{S} \subseteq \mcl{M}\setminus \{1\} \right\}. \nonumber %\label{Rt_region}
\end{align}
Hence, the underlying optimization problem has the following form
\begin{align}
\min_{\mbf{R}} \sum_{i\in \mcl{M} \setminus \{1\}} \alpha_i R_i,~~\text{s.t.}~\mbf{R} \in \mcl{R}_1. \label{one_user_opt}
\end{align}
Optimization problem~\eqref{one_user_opt} can be solved analytically due to the fact that the set function
\begin{align}
f(\mcl{S})=H(X_{\mcl{S}}|X_{\mcl{S}^c},X_1),~~\forall \mcl{S}\subseteq \mcl{M} \setminus \{1\}
\end{align}
is supermodular (see~\cite{F05} for the formal definition). Therefore, optimization problem~\eqref{one_user_opt} is over a supermodular polyhedron $\mcl{R}_1$. From the combinatorial optimization theory it is known that Edmonds' greedy algorithm~\cite{E70} renders
an analytical solution to this problem (see Algorithm~\ref{alg:edm}).
%Before we solve the optimization problem~\eqref{one_user_opt} let us introduce some concepts from
%combinatorial optimization theory. A set function $f:2^{\mcl{M} \setminus \{1\}} \rightarrow \mathbb{R}$ is supermodular if
%\begin{align}
%f(\mcl{S})+f(\mcl{T})\leq f(\mcl{S} \cup \mcl{T}) + f(\mcl{S} \cap \mcl{T}),~\forall \mcl{S},\mcl{T}\ \subseteq \mcl{M} \setminus \{1\}. \nonumber
%\end{align}
%It is not hard to show that the function
%\begin{align}
%f(\mcl{S})=H(X_{\mcl{S}}|X_{\mcl{S}^c},X_1),~~\forall \mcl{S}\subseteq \mcl{M} \setminus \{1\}
%\end{align}
%is supermodular.
%Therefore, the optimization problem~\eqref{one_user_opt} is over a supermodular polyhedron $\mcl{R}_1$.
%From the combinatorial optimization theory it is known that Edmonds' greedy algorithm~\cite{E70} renders
%an analytical solution to this problem (see Algorithm~\ref{alg:edm}).
\begin{algorithm}
\caption{Edmonds' algorithm applied to our problem}
\label{alg:edm}
\begin{algorithmic}[1]
\STATE Set $j_1,j_2,\ldots,j_{m-1}$ to be an ordering of $\{1,2,\ldots,m\} \setminus \{1\}$ such that $\alpha_{j_1}\leq \alpha_{j_2}\leq \cdots \leq \alpha_{j_{m-1}}$.
\FOR {$i=1$ to $m-1$}
\STATE $R^{*}_{j_i}=H(X_{j_i}|X_{t_1},X_{j_1},X_{j_2},\ldots,X_{j_{i-1}})$.
\ENDFOR
\end{algorithmic}
\end{algorithm}

\begin{example}\label{exp1}
Consider a system with $m=6$ terminals $\mcl{M}=\{1,2,3,4,5,6\}$.
For convenience, we express the underlying data vector as
$\mathbf{W}=\left[
                                               \begin{array}{ccc}
                                                 a & b & c  \\
                                               \end{array}
                                             \right]^T \in \mathbb{F}^3_{q^n}$,
where $a, b, c$ are independent uniform random variables in ${\mathbb F}_{q^n}$.
Let us consider the case where each node has the following observations:
$\mbf{X}_1=\{a+b\}$, $\mbf{X}_2=\{a+c\}$, $\mbf{X}_3=\{b+c\}$, $\mbf{X}_4=\{a\}$, $\mbf{X}_5=\{b\}$, $\mbf{X}_6=\{c\}$.
Let us assume that user $1$ is interested in recovering the vector
$\mbf{W}$ such that underlying communication cost is $\sum_{i=2}^6 R_i$.

It immediately follows from Algorithm~\ref{alg:edm} that a solution to this problem
is $R^{*}_4=R^{*}_6=1$, and $R^{*}_2=R^{*}_3=R^{*}_5=0$. In other words, user~$1$ is missing $2$ linear equations in order to be able
to decode all $3$ data packets.
\end{example}

\subsection{Deterministic Algorithm when $|\mcl{A}|>1$}

In this section we extend the results from the previous section to the case where the set $\mcl{A}$
contains arbitrary number of users.
Optimization problem~\eqref{problem1} can be written as follows

\begin{align}
&\min_{\mbf{Z},\mbf{R}} \sum_{i=1}^m \alpha_i Z_i, \label{problem2}  \\
&~~~~~~\text{s.t.}~Z_i \geq R_i^{(t_l)},~~\forall l \in \mcl{A},~\forall i \in \mcl{M}\setminus \{l\}, \nonumber \\
&~~~~~~~~~~\mbf{R}^{(l)} \in \mcl{R}_l,~~\forall l \in \mcl{A}, \nonumber
\end{align}
where
\begin{align}
\mcl{R}_l = \left\{ \mbf{R} : R(\mcl{S})\geq H(X_{\mcl{S}}|X_{\mcl{S}^c},X_1),~\forall \mcl{S} \subseteq \mcl{M}\setminus \{l\} \right\}. \nonumber
\end{align}
Equivalence between the optimization problems~\eqref{problem1}
and \eqref{problem2} follows from the fact that transmissions of all terminals in $\mcl{M}$ have to be such that
all users in $\mcl{A}$ can learn $X_{\mcl{M}}$. Optimization problem~\eqref{problem2} has an exponential number constraints, which makes it challenging to solve in polynomial time. To obtain a polynomial time solution we consider the Lagrangian dual of problem~\eqref{problem2}.

\begin{align}
&\max_{\mbf{\Lambda}} \sum_{l=1}^k g^{(l)} (\mbf{\Lambda}^{(l)}), \label{dual} \\
&~\text{s.t.}~\sum_{l=1}^k \lambda_i^{(l)}=\alpha_i,~\lambda_i^{(l)}\geq 0,~~\forall l \in \mcl{A},~\forall i \in \mcl{M} \setminus \{l\}, \nonumber
\end{align}
where
\begin{align}
g^{(l)}(\mbf{\Lambda}^{(l)}) = \min_{\mbf{R}^{(l)}} \sum_{i \in \mcl{M} \setminus \{l\}} \lambda_i^{(l)} R_i^{(l)},~~\text{s.t.}~~\mbf{R}^{(l)}\in \mcl{R}_l. \label{problem3}
\end{align}
Dual variable $\mbf{\Lambda}$ in the above problem is represented in matrix form as follows.
\begin{align}
\mbf{\Lambda}=\left[
                \begin{array}{cccc}
                  \lambda_1^{(1)} & \lambda_2^{(1)} & \cdots & \lambda_m^{(1)} \\
                  \lambda_1^{(2)} & \lambda_2^{(2)} & \cdots & \lambda_m^{(2)} \\
                  \vdots & \vdots & \ddot & \vdots \\
                  \lambda_1^{(k)} & \lambda_2^{(k)} & \cdots & \lambda_m^{(k)} \\
                \end{array}
              \right]. \label{lmb}
\end{align}
We denote by $\mbf{\Lambda}_i$ and $\mbf{\Lambda}^{(l)}$, the $i^{th}$ column vector and $l^{th}$
row vector of the matrix $\mbf{\Lambda}$, respectively. Moreover, we denote by
\begin{align}
\mbf{\tilde{R}}=
\left[
  \begin{array}{cccc}
    R_1^{(1)} & R_2^{(1)} & \cdots & R_m^{(1)} \\
    R_1^{(2)} & R_2^{(2)} & \cdots & R_m^{(2)} \\
     \vdots & \vdots & \ddot & \vdots \\
    R_1^{(k)} & R_2^{(k)} & \cdots & R_m^{(k)} \\
  \end{array}
\right] \label{rate_matrix}
\end{align}
the rate matrix whose $l^{th}$ row, here denoted by $\mbf{\tilde{R}}^{(l)}$, represents an optimizer of the problem~\eqref{problem3} w.r.t. the weight vector $\mbf{\Lambda}^{(l)}$. In order to ensure consistency with the optimization problem~\eqref{dual} observe that
$\lambda_{l}^{(l)}=0$, and $R_{l}^{(l)}=0$, $\forall l=1,\ldots,k$.

For any given user $l \in \mcl{A}$, the objective function~\eqref{problem3} of the dual problem~\eqref{dual} can be computed analytically
using Algorithm~\ref{alg:edm}. The optimization problem~\eqref{dual} is a linear program (LP) with $\mcl{O}(m \cdot k)$
number of constraints, which makes it possible to solve it in polynomial time (w.r.t. number of terminals).
To solve the optimization problem~\eqref{dual} we apply a subgradient method, as described below.

Starting with a feasible iterate $\mbf{\Lambda}[0]$ w.r.t. the optimization problem~\eqref{dual},
every subsequent iterate $\mbf{\Lambda}[n]$ can be recursively represented as an Euclidian projection of the vector
\begin{align}
\mbf{\Lambda}_i[n] = \mbf{\Lambda}_i[n-1] + \theta [n-1]\cdot \mbf{\tilde{R}}_i[n-1],~~\forall i \in \mcl{M} \label{iterate}
\end{align}
onto the hyperplane $\left\{ \mbf{\Lambda}_i\geq \mbf{0} | \sum_{l=1}^k \lambda_i^{(l)} = \alpha_i \right\}$,
where $\mbf{\tilde{R}}_i[n-1]$ is the $i^{th}$ column of the rate matrix $\mbf{\tilde{R}}[n-1]$.
The Euclidian projection ensures that every iterate $\mbf{\Lambda}[n]$ is feasible w.r.t. the optimization problem~\eqref{dual}.
It is not hard to verify that the following initial choice of $\mbf{\Lambda}[0]$ is feasible w.r.t. the problem~\eqref{dual}.
\begin{align}
\lambda_i^{(l)}[0]=
\begin{cases}
\frac{\alpha_i}{k}  &  \text{if}~i \not \in \mcl{A} \\
\frac{\alpha_i}{k-1}  &  \text{if}~i \in \mcl{A} \setminus \{l\} \label{lambda_initial} \\
0               &  \text{if}~i=l
\end{cases},~~\forall i \in \mcl{M},~\forall l \in \mcl{A}.
\end{align}

By appropriately choosing the step size $\theta[n]$ in each iteration~\eqref{iterate}, it is guaranteed that the subgradient method
described above converges to the optimal solution of the problem~\eqref{dual}. To recover the primal optimal solution
from the iterates $\mbf{\Lambda}[n]$ we use results from~\cite{sherali1996recovery}, where at each iteration of~\eqref{iterate},
the primal iterate is constructed as follows.
\begin{align}
\mbf{\hat{R}}[n] = \sum_{j=1}^n \mu_j^{(n)} \mbf{\tilde{R}}[j], \label{pr_recovery}
\end{align}
where
\begin{align}
\sum_{j=1}^n \mu_j^{(n)}=1,~\mu_j^{(n)}\geq 0,~\text{for}~j=1,2,\ldots,n.
\end{align}
By carefully choosing the step size $\theta[n]$, $\forall n$ in \eqref{iterate} and the convex combination coefficients $\mu_j^{(n)}$,
$\forall j=1,\ldots,n$, $\forall n$,
it is guaranteed that \eqref{pr_recovery} converges to the minimizer of \eqref{problem2}, and therefore to the minimizer
of the original problem~\eqref{problem1}. In~\cite{sherali1996recovery}, the authors proposed several choices for $\{\theta[n]\}$ and
$\{\mu_j^{(n)}\}$ which lead to the primal recovery. Here we list some of them.
\begin{enumerate}
\item $\theta[n]=\frac{a}{b+cn}$, $\forall n$, where $a>0$, $b\geq 0$, $c>0$,  \\
      $\mu_j^{(n)}=\frac{1}{n}$, $\forall j=1,\ldots,n$, $\forall n$,
\item $\theta[n]=n^{-a}$, $\forall n$, where $0<a<1$, \\
      $\mu_j^{(n)}=\frac{1}{n}$, $\forall j=1,\ldots,n$, $\forall n$.
\end{enumerate}

Now, it is only left to compute an optimal rate allocation w.r.t to the problem defined in~\eqref{problem1}.
Let $\mbf{R}^{*}$ and $\mbf{Z}^{*}$ be the optimal rate vectors of the problems~\eqref{problem1} and~\eqref{problem2}, respectively.
As we pointed out earlier $\mbf{R}^{*}=\mbf{Z}^{*}$, where $\mbf{Z}^{*}$ can be computed from the matrix $\mbf{\hat{R}}[n]$
for a sufficiently large $n$, as follows
\begin{align}
Z_i^{*} = \max \left\{ \hat{R}_i^{(1)}[n], \hat{R}_i^{(2)}[n], \ldots, \hat{R}_i^{(k)}[n] \right\},~~\forall i \in \mcl{M}.
\end{align}
Pseudo code of the algorithm described in this section is shown below (see Algorithm~\ref{alg:optimal}).
\begin{algorithm}
\caption{Optimal \emph{DE} rate allocation}
\label{alg:optimal}
\begin{algorithmic}[1]
\STATE Initialize $\mbf{\Lambda}[0]$ according to~\eqref{lambda_initial}
%\FOR {$i=1$ to $m$}
%\FOR {$l=1$ to $k$}
%\STATE $\lambda_i^{(l)}[0]=
%        \begin{cases}
%        \frac{\alpha_i}{k}  &  \text{if}~i \in \{k+1,\ldots,m\} \nonumber \\
%        \frac{\alpha_i}{k-1}  &  \text{if}~i \in \{1,\ldots k\} \setminus \{l\} \nonumber \\
%         0               &  \text{if}~i=l \nonumber
%        \end{cases}$
%\ENDFOR
%\ENDFOR
\STATE Set $\theta[n]=\frac{1}{n+1}$, $\forall n$,~~$\mu_j^{(n)}=\frac{1}{n}$,~$\forall j=\{1,\ldots,n\}$
\FOR {$n=1$ to $\bar{n}$}
\FOR {$l=1$ to $k$}
\STATE Compute $\tilde{\mbf{R}}^{(l)}[n]$ using Algorithm~\ref{alg:edm} for the weight vector $\mbf{\Lambda}^{(l)}[n]$
\ENDFOR
\STATE Project $\mbf{\Lambda}_i[n] = \mbf{\Lambda}_i[n-1] + \theta [n-1]\cdot \mbf{\tilde{R}}_i[n-1]$
       onto the hyperplane $\left\{ \mbf{\Lambda}_i\geq \mbf{0} | \sum_{l=1}^k \lambda_i^{(l)} = \alpha_i \right\}$.
\ENDFOR
\STATE $\mbf{\hat{R}}[\bar{n}] = \sum_{j=1}^{\bar{n}} \mu_j^{(\bar{n})} \mbf{\tilde{R}}[j]$
\STATE $R_i^{*} = \max \left\{ \hat{R}_i^{(1)}[\bar{n}], \hat{R}_i^{(2)}[\bar{n}], \ldots, \hat{R}_i^{(k)}[\bar{n}] \right\}$
\end{algorithmic}
\end{algorithm}

\subsection{Code Construction for the Linear Source Model}

In this Section we briefly address the question of the optimal code construction for the linear source model.
For that matter, let us consider the following example.
\begin{example}
Let us consider the same source model as in Example~\ref{exp1}, where $\mcl{A}=\{1,2,3\}$,
and the objective function is $\sum_{i=1}^6 R_i$. Applying the algorithm described above, we obtain
\begin{align}
R^{*}_1=R^{*}_2=R^{*}_3=\frac{1}{4},~~R^{*}_4=R^{*}_5=R^{*}_6=\frac{1}{2}. \label{chan}
\end{align}
\end{example}

\begin{figure}[h]
\begin{center}
\includegraphics[scale=0.47]{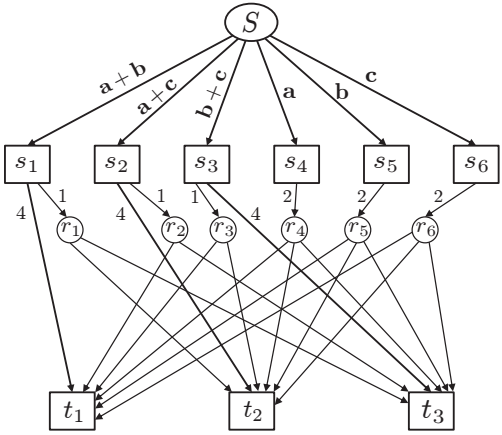}
\end{center}
\vspace{-0.1in}
\caption{Multicast network constructed from the source model and the optimal rate tuple
         $R^{*}_1=R^{*}_2=R^{*}_3=2$, $R^{*}_4=R^{*}_5=R^{*}_6=1$. Each user receives side-information from
         ``itself'' through links $(s_i,r_i)$, $i=1,2,3$, and from the other terminals
         through links $(t_i,r_j)$, $i=1,\ldots,6$, $j=1,2,3$, $i\neq j$.}\label{fig:multicast}
\end{figure}

This solution suggests that in order to design a scheme that performs optimally, it is necessary to split
all the packets into $4$ equally sized chunks. In other words, terminals' observations can be written as
$\mbf{X}_1=\mbf{a}+\mbf{b}=\{a_1+b_1,a_2+b_2,a_3+b_3,a_4+b_4\}$, $\mbf{X}_2=\mbf{a}+\mbf{c}=\{a_1+c_1,a_2+c_2,a_3+c_3,a_4+c_4\}$, etc.,
where all $a_i$'s, $b_i$'s and $c_i$'s belong to $\mathbb{F}_{q^{n/4}}$. For this ``extended'' source
model we have that the optimal rate allocation is $R^{*}_1=R^{*}_2=R^{*}_3=1$, $R^{*}_4=R^{*}_5=R^{*}_6=2$.

Next question we need to address is how to design transmissions of each user? Starting
from an optimal (integer) rate allocation, we first construct the corresponding multicast network (see Figure~\ref{fig:multicast}).
In this construction, notice that there are several types of nodes. First, there is a super node $S$ that possesses all the packets.
Each user in the set $\mcl{A}$ plays the role of a transmitter and a receiver, while the helpers act only as transmitters.
To model this, we denote $s_1,\ldots,s_6$ to be the ``sending'' nodes, and $r_1$, $r_2$ and $r_3$ to be the receiving nodes.
To model the side-information at users $1$, $2$ and $3$, we introduce links $(s_i,r_i)$, $i=1,2,3$, of capacity $4$, which are
routing the users' observations to the corresponding receiving nodes. To model the broadcast nature of each transmission, we introduce
``dummy'' nodes $t_1,\ldots,t_6$, such that the capacity of the links $(s_i,t_i)$ is the same as link capacity $(t_i,r_j)$, $j\neq i$,
and is equal to $R^{*}_i$, $\forall i \in \mcl{M}$.

To solve for actual transmissions of each terminal, we apply the algebraic network coding approach~\cite{KM03},
with appropriately designed source matrix $\mbf{A}$ which corresponds to the side-information of all terminals.
%where the source matrix $\mbf{A}$ is given by
%\begin{align}
%\mbf{A}=\left[
%          \begin{array}{cccc}
%            \mbf{A}_1^T & \ldots & \mbf{A}_m^T & \mbf{0} \ldots \mbf{0} \\
%          \end{array}
%        \right].
%\end{align}
Finally, the network code for the data exchange problem with helpers can be constructed in polynomial time from the algorithms
provided in~\cite{H05} which are based on a simultaneous transfer matrix completion.

%\section{Extensions to Other Omniscience Problems}
\section{Conclusion and Extensions}
In this paper we study the data exchange problem with helpers. We provide a deterministic polynomial time algorithm for minimizing the weighted sum-rate cost of communication.
We show that the data exchange problem with only one user and many helpers can be solved analytically using Edmonds' algorithm. Further using single user solution as a building block we show how one can solve the more general problem with arbitrary number of users.
%Algorithm~\ref{alg:optimal} provides a general tool for solving the communication problems in polynomial time, when a group
%of users is interested in obtaining the same information, and when it is possible to solve a single user case in polynomial time.
Several extensions are of interest.
For instance, we can consider a modification of the original data exchange problem where only helpers are allowed to transmit.
Starting from a single user case, it is easy to see that an achievable rate tuple must satisfy
all the cut-set constraints over the helper set such that the user is always on the receiving side of the cut.
Minimizing the weighted sum-rate cost over all achievable rate tuples can again be done using Edmonds' algorithm
(see Algorithm~\ref{alg:edm}). Finally, extension to the multiple user case corresponds to the weighted sum-rate minimization over all
rate tuples that are simultaneously achievable for all users. This optimization problem can be solved in polynomial time
using the same approach as in Algorithm~\ref{alg:optimal}.

%When $\mcl{A}=\{1\}$, the underlying optimization problem can be written as follows:
%\begin{align}
%&\min_{\mbf{R}} \sum_{i \in \mcl{M} \setminus \mcl{A}} \alpha_i R_i, \label{problem4} \\
%&~~~~~~\text{s.t.} R(\mcl{S})\geq H(X_{\mcl{S}}|X_{\mcl{S}^c},X_1),~\forall \mcl{S} \subseteq \mcl{M} \setminus \mcl{A}. \nonumber
%\end{align}
%Problem~\eqref{problem4} can be analytically solved by using Algorithm~\ref{alg:edm} applied to the helper set $\mcl{M} \setminus \mcl{A}$.
%Hence, by using Algorithm~\ref{alg:optimal} one can solve the optimization problem with arbitrary number of users in $\mcl{A}$
%\begin{align}
%&\min_{\mbf{R}} \sum_{i=1}^m \alpha_i R_i, \label{problem5} \\
%&~~~~~\text{s.t.}~~R(\mcl{S})\geq H(X_{\mcl{S}}|X_{\mcl{S}^c},X_l),~\forall \mcl{S} \subseteq \mcl{M}\setminus \mcl{A},~\forall l \in \mcl{A}. \nonumber
%\end{align}
%in polynomial time.

%\input{linear_example.tex}
%\input{det_algorithm.tex} \label{sec:alg}
%\input{Intro.tex}
%\input{Model.tex}
%\input{linear_example.tex}
%\input{det_algorithm.tex} \label{sec:alg}
%\input{Conclusion.tex} \label{sec:conclusion}
%\input{appendix.tex} \label{sec:app}
%\input{Connection_to_KeyAgreement.tex}
%\input{Deterministic.tex}
%\input{Appendix.tex}

%\bibliographystyle{IEEEtran}
%%\footnotesize{
%\bibliography{DatExc}

\end{document}